# Adaptive Full Duplex Communications in Cognitive Radio Networks

Vahid Towhidlou
Centre for Telecommunication Research
King's College London

Mohammad Shikh Bahaei
Centre for Telecommunication Research
King's College London

*Abstract*— In this paper we propose a novel adaptive scheme for full duplex communication of secondary users (SUs) in a cognitive radio network. The secondary network operates in three modes; Cooperative Sensing (CS), Full Duplex Transmit and Sensing (FDTS), and Full Duplex Transmit and Receive (FDTR). In the CS mode, the secondary nodes detect the activity of primary users (PUs) through a novel cooperative MAC protocol and will decide the system's mode of operation in the subsequent spectrum hole. In the FDTS mode one of the SUs senses the PUs' activity continuously whilst transmitting to another node. In the FDTR mode, the SUs would communicate bidirectionally in an asynchronous full duplex (FD) manner, with decreased maximum and average collision durations. Analytical closed forms for probability of collision, average collision duration and cumulative collision duration, as well as throughput of the SU network are derived, and performance of the proposed protocol in terms of above-mentioned metrics, its effectiveness, and advantages over conventional methods of sensing and transmission are verified via simulations.

## I. INTRODUCTION

The unprecedented growth in wireless devices and mobile data traffic have raised an earnest attention towards finding new solutions for more efficient utilization of the wireless spectrum. Cognitive radio (CR) and full duplex (FD) communications are two promising technologies recently developed to enhance spectrum utilization and network efficiency, and combination thereof will improve the performance even further. On the other hand, abundant work exists on adaptive transmission methods across physical and MAC layers (e.g. see [12-16]) both in standard and cognitive networks.

In general, a CR network (CRN) is non time slotted which means PUs can be active or inactive at any time during the whole secondary frame duration. Traditional spectrum sensing schemes, such as Listen Before Talk (LBT) in non time slotted networks, usually do not guarantee good protection for the PUs, as they may become active at anytime during SUs' transmission, and this will cause collision. Some studies [1-4] have proposed FD spectrum sensing to alleviate this problem. In such works the transmitting SU keeps sensing the presence of PUs constantly during the transmission slot, and upon detection of a PU signal, will stop transmission to avoid collision [1]. Although such techniques will noticeably improve PU protection, but will not enhance throughput of the CR network. The authors in [5] presented a threshold-based sensing-transmission scheme in order to maximize SU throughput with minimum collision with PUs in a half duplex (HD) network, which may be applied in FD mode as well. Although this scheme shows enhancement over traditional LBT schemes in terms of throughput and collision probability in HD mode, but in FD mode it would not be optimal. Inspired by that idea, the authors in [6] proposed a similar approach for a FD scenario in a CRN with four modes of operation; sensing-only, transmit and sensing, transmit and receive, and channel selection, and developed an optimal mode-selection strategy to maximize the SU throughput for a given PU collision probability. However, the average and maximum collision durations were not optimized, and it could be very long if the primary SNR at secondary receivers is weak.

In [7], we considered an interweave CR system and introduced the CS and FDTR modes of operation in which SUs were capable of partial self-interference suppression (SIS). The proposed system exploited cooperative sensing and bidirectional FD communication of cognitive network. The asynchronous transmission of secondary nodes could reduce collision durations, although the system could experience long collision durations in FDTR mode if the primary's signal was not strong enough to be detected in FDTR mode.

In order to alleviate the shortcomings of the system proposed in [7], in this paper we have presented a novel adaptive scheme for full duplex CRNs based on three modes of operation; cooperative sensing (CS), full duplex transmit and sensing (FDTS), and full duplex transmit and receive (FDTR). The proposed scheme not only enhances the CRN's throughput, but will decrease the average and maximum collision durations with primary signal while avoiding long or endless collision duration in varying channel conditions.

In the CS mode when a PU is active, SUs keep sensing the channel through consecutive sensing intervals in a cooperative manner. This will increase detection probability and there would be no collision with primary network during its busy period. In this mode, the system decides on the network's mode of operation for the following idle period of primary network, based on the existing conditions of PU-SU channels. If the system decides that the whole network is in a condition that bidirectional FD communication is feasible with tolerable inflictions, it will switch to FDTR mode upon disappearance of PU signal. However, if the network status is not reliable enough for such mode, then the system will prefer to switch to the conservative FDTS mode in which one of the SUs detects the return of a primary signal continuously while transmitting to the other node, in a full duplex manner.

In the FDTR mode, the cognitive system is notified of the return of a primary signal through collision. In order to minimize the interferences impinged by secondary on primary, we have considered asynchronous transmission of SUs in this mode. It is shown that such alteration in transmission timing in the secondary network will decrease collision duration to half or less compared to that in the traditional sensing methods.

The main contributions of this paper are as follows. We have presented a novel adaptive scheme for FD operation in CRNs which consists of three modes of operation. Mode selection based on a dual-threshold criterion and the cooperative MAC protocol in the CS mode are proposed for the first time. The impact of asynchronous transmission on network metrics such as probability of false alarm and detection, collision probability and duration, and the CR network average throughput are mathematically analyzed and it is shown that our scheme outperforms the conventional methods in terms of CRN metrics.

The rest of this paper is organized as follows. Section II describes the system model and protocol description. In Section III we derive exact closed form expressions for probability of detection and false alarm in CS and FDTR modes, probability of collision and probability density function of collision duration, and the average CR network throughput for synchronous and asynchronous modes, and the traditional LBT scheme for comparison. Numerical results are provided in Section IV, and conclusion is presented in Section V.

## II. SYSTEM MODEL

We consider an interweave CRN consisting of multiple PUs and one pair of SUs, which opportunistically access the licensed PUs' channel as shown in Fig. 1. We focus on a bidirectional channel, and assume all channels between primary to secondary, and secondary to secondary nodes are Rayleigh block fading. PUs may become active or inactive at any time, modeled as an alternating ON/OFF continuous-time Markov process [8]. The ON and OFF times are Exponentially distributed with mean durations of $\mu^{-1}$ and $\lambda^{-1}$ respectively. SUs are capable of operating in the FD mode. This requires each SU be capable of perfect or partial SIS. Here we do not consider SIS methods and just quantify this capability by SIS factor $\beta$ of the SUs. SIS factor is the ratio between the residual self-interference (SI) signal and the original one. In partial SIS only the fraction $1 - \beta$ of the original SI signal is cancelled.

The secondary network operates in three modes; Cooperative Sensing (CS), Full Duplex Transmit and Sensing (FDTS), and Full Duplex Transmit and Receive (FDTR).

When the primary network is active, the system operates in the CS mode. In this mode, both SUs keep listening to the channel during continuous sensing slots in tandem, and do not transmit. Each sensing slot is $T_s$ seconds which is a fraction of transmission frame duration $T$. In contrast with traditional sensing methods based on energy detection with one threshold, here we set up two threshold values; $\epsilon_0$ and $\epsilon_1$ where $\epsilon_0 < \epsilon_1$. The lower threshold ($\epsilon_0$) is used for detecting the presence of a primary signal, and the higher threshold ($\epsilon_1$) is employed for selecting the mode of operation of CRN in the subsequent OFF period. Cooperation between SUs in sensing the primary's presence increases the detection probability, which improves protection for the primary network.

### Dual-Threshold Detection (DTD)

In some cases, where the PU signal at SUs' receivers is weak (e.g. when the SUs are far from a PU transmitter but close to the PU receiver), the effect of a weak PU signal on SUs' transmissions is not that much to cause errors in SUs' transmission. Hence the SUs' will not notice the return of a PU signal through collision event and may continue FD communication and impinge long interference on primary's signal. To avoid such incidences, the CRN operates in the FDTS mode. For this purpose, we introduce dual threshold detection scheme with a pre-defined mode selection threshold ($\epsilon_1$) and primary presence threshold ($\epsilon_0$).

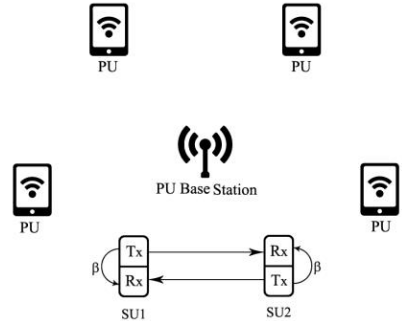

Fig. 1. System model of our network

In each sensing slot, each SU first compares the detected energy ($\mathcal{E}_d$) with $\epsilon_1$. If $\mathcal{E}_d > \epsilon_1$, it means that PU is present and its power is high enough for FDTR mode operation in the subsequent OFF period. However if $\mathcal{E}_d < \epsilon_1$, then it is compared with the lower threshold. $\epsilon_0 < \mathcal{E}_d < \epsilon_1$ means that a primary signal with a low energy level is present. Hence, for a conservative system avoiding long collision periods, the CR system should operate in the FDTS mode, during the next upcoming OFF period of primary network. And $\mathcal{E}_d < \epsilon_0$ is interpreted that the primary signal is not present anymore and there is an opportunity for SUs to start communication in the suitable mode decided within previous sensing slots.

### Cooperative Sensing and Mode selection MAC

Cooperative sensing is carried out through a handshaking MAC protocol. When one SU finds the channel not in use (i.e. $\mathcal{E}_d < \epsilon_0$), it transmits ready-to-send (RTS) signals via a report channel in the next sensing slot and at the same time listens to receive a RTS signal back from the peer SU. We have two types of RTS signals; RTS1 and RTS2, which are very short and assumed to be error free. RTS1 is declared for operation in FDTS mode, and RTS2 for FDTR mode. If both SUs declare RTS1 signals, they will operate in FDTS mode and one of the nodes by default will be the transmitting and sensing node, and the other one the destination. But if one SU emits an RTS2 signal and the other one emits RTS1 or RTS2, then both SUs will enter into bidirectional FDTR mode. And finally, if only one SU sends RTS signals but does not receive an RTS back, it means that the other end has not detected the channel idle, so the sensing continues in the following sensing intervals until both arrive at the same decision.

In the FDTS mode, one of the SUs acts as the source, and the other one as destination. The source unit transmits to the destination, and at the same time keeps sensing the channel for the return of primary signal. In this mode, communication between SUs is in half duplex (HD) mode, but the source node operates in full duplex transmit and sensing manner.

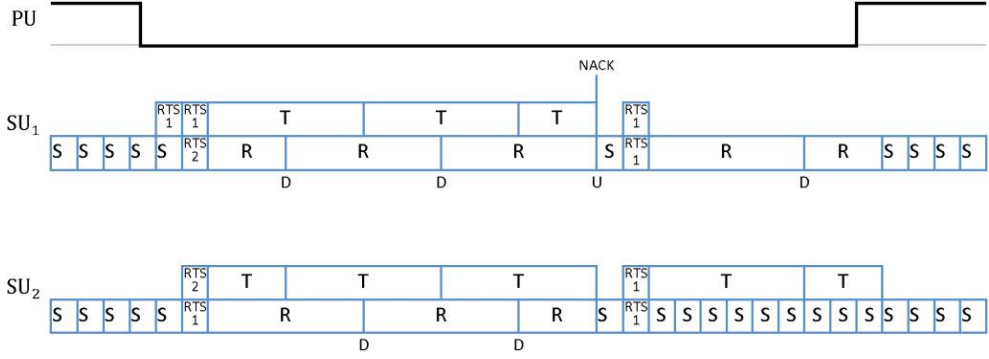

Fig. 2. CS, FDTR and FDTS modes and switching between them through a MAC handshaking protocol.

In the FDTR mode, both SUs transmit and receive their data in FD manner over the same channel as long as no primary signal has returned to the channel. When primary reappears, the primary and secondary signals collide and this will result in errors in secondary network communication which will indirectly inform the SUs of the presence of a PU and will force them to cease transmission instantly and switch back to the CS mode. In this mode there is no explicit sensing of PUs' activities, and any SU that cannot decode the received packet or frame without error (Undecode event), or receives a NACK from the other SU (NACK event), would interpret it as a result of PU return. The whole process is shown in Fig. 2.

Although operating in the FDTR mode enhances the SUs' throughput, the SUs will not be able to monitor the PU's state within a transmission frame. Hence, the probability of colliding with PUs will be higher than in traditional methods of sensing such as Listen Before Talk (LBT). In addition, the PU's signal may affect SUs differently due to different fading or interference conditions of PU-SU channels. It means that a SU transmission may not be affected by PU signal, due to deep fading of the PU-SU channel, and no collision would be declared although PU is on. This may increase collision probability as well. But such conditions may happen in traditional methods too. In our scheme there may be errors not caused by collision with a primary signal. Fading, noise or any other interferences may cause such errors which will in turn cause false alarms, interrupt SUs' transmissions, and decrease SU throughput. However, our method is a conservative way to avoid any probable collision with PUs to guarantees high primary's protection. On the other hand, duration of sensing slots is small compared to the secondary frames, and CS decreases the probability of consecutive false alarms. Therefore, the throughput degradation due to general packet or frame errors is small, as will be seen in the sequel.

The indirect sensing method applied in the FDTR mode which is based on collision event usually results in long collision durations and may not be acceptable by the primary network. In order to alleviate this, we have considered asynchronous transmissions as proposed in [7]. This will decrease the average collision duration. For the case of equal transmission frames, the optimum timing difference would be $T/2$ for shortest average collision duration.

## III. PERFORMANCE ANALYSIS

### A. Sensing Metrics

If prior knowledge of the PU signal is unknown, the energy detection method is optimal for detecting zero-mean constellation signals [9]. For circularly symmetric complex Gaussian signals, the probability of false alarm and detection of each node are as follows, respectively [9]:

$$P_f(\epsilon, T_s) = Q\left(\left(\frac{\epsilon}{\sigma_u^2} - 1\right)\sqrt{f_s \cdot T_s}\right), \quad (1)$$

$$P_d(\epsilon, T_s) = Q\left(\left(\frac{\epsilon}{\sigma_u^2} - \gamma - 1\right)\sqrt{\frac{f_s \cdot T_s}{2\gamma + 1}}\right), \quad (2)$$

where $\epsilon$ is the detection threshold, $\sigma_u^2$ is the noise variance, $\gamma$ is primary's SNR at secondary receiver, $f_s$ is sampling rate, and $Q(\cdot)$ is the complementary error function.

For a lower probability of collision, we will fuse the results of two sensing units in the CS mode and the hypothesis is that PU is in off state when both SUs have sensed the channel free. This will decrease misdetection probability to a very low value but will increase the probability of false alarm, as follows:

$$P_{f1} = 2P_f(\epsilon, T_s) - P_f(\epsilon, T_s)^2, \ P_{d1} = 2P_d(\epsilon, T_s) - P_d(\epsilon, T_s)^2 \quad (3)$$

$P_{f1}$ and $P_{d1}$ are probabilities of false alarm and detection in the CS mode. In FDTS mode, there is no cooperation in sensing and the residual self-interference signal will affect the sensing probability which are analyzed thoroughly in [2] and [3] and are not reproduced here. In the FDTR mode detection method is different and is based on NACK and Undecode events when a collision with primary happens. Since we assume that transmission errors in NACKs are negligible, the probability of not receiving a NACK (misdetection) despite the presence of PU, is the probability that both channels from PU to SU1 and SU2 be in deep fade and the reverse channels not in deep fade, which is very small and may be ignored.

On the other hand, false alarm happens when an error in a frame triggers a NACK due to reasons other than collision with the primary. Hence within the FDTR mode, false alarm probability ($P_{f2}$) is the same as average Frame Error Rate ($\overline{FER}$) [1]. We assume that both SUs have the same average frame error

---
[1] If NACK and Undecode events are based on packet errors instead of frame errors, then $P_{f2} = \overline{PER}$.

rate. Average frame error rate for transmissions with $N_f$ packets in a frame would be:

$$\overline{FER} = 1-(1-\overline{PER})^{N_f}. \tag{4}$$

$\overline{PER}$ is the average packet error rate of the secondary nodes. Here we use the approximate value for instantaneous packet error rate given in [10] which is widely used in many works:

$$PER(\gamma) = \begin{cases} 1, & 0 < \gamma < \gamma_t \\ \alpha \exp(-g\gamma), & \gamma \geq \gamma_t \end{cases}. \tag{5}$$

where $\gamma$ is the instantaneous signal-to-noise ratio and $(\alpha, g, \gamma_t)$ are mode dependent parameters found by least-squares fitting to the exact packet error rate. In a Rayleigh fading channel, $\gamma$ follows an Exponential distribution. For an SU with imperfect SIS where the instantaneous SNR is set to $\gamma_{S_1}/(1 + \beta\gamma_{S_2})$ in which $\gamma_{S_1}$ is the SNR of SU1 at SU2 and $\beta\gamma_{S_2}$ is the residual self-interference of SU2 after partial cancellation, the average value of (1) would be:

$$\overline{PER} = \mathbb{E}[PER(\gamma)] = \int_0^\infty PER\left(\frac{\gamma_{S_1}}{1 + \beta\gamma_{S_2}}\right) f_{S_1}(\gamma_{S_1}) d\gamma_{S_1}$$
$$= 1 - \frac{g\bar{\gamma}_{S_1}}{1 + \beta\bar{\gamma}_{S_2} + g\bar{\gamma}_{S_1}} \exp\left(\frac{-(1 + \beta\bar{\gamma}_{S_2})\gamma_t}{\bar{\gamma}_{S_1}}\right). \tag{6}$$

### B. Collision Probability and Collision Duration Distribution

Collision with primary signal may occur at the beginning or the end of a secondary frame. Pre-collision happens when the network is in the CS mode and misdetection occurs, and post-collision occurs within the FDTR mode. Therefore, probability of collision, $P_{Async}^{col}$ in our scheme would be

$$P_{Async}^{col} = P(\mathcal{H}_0)\left(1 - P_{f_2}\frac{T_s}{T}\right)P_\tau + P(\mathcal{H}_1)(1 - P_{d1}), \tag{7}$$

where $P_\tau = 1 - e^{-\lambda T}$ is the probability that primary signal reappears within a SU frame of duration $T$. In the traditional LBT scheme, probability of collision can be written as:

$$P_{LBT}^{col} = P(\mathcal{H}_0)(1 - P_{f1})P_\tau + P(\mathcal{H}_1)(1 - P_{d1}). \tag{8}$$

$\mathcal{H}_0$ is the event that the PU is not active at the start of the frame, and its complement is $\mathcal{H}_1$. In our system where primary's idle and busy times are exponential with parameters $\lambda$ and $\mu$ respectively, the respective *a priori* probabilities of primary's activity at the start of each secondary frame are $P(\mathcal{H}_0) = \mu/(\lambda + \mu)$ and $P(\mathcal{H}_1) = \lambda/(\lambda + \mu)$.

When a collision occurs in FDTR mode, it would last until the first NACK signal is declared from any of the SUs. Having assumed error-free NACK signals, in asynchronous transmission mode when there is a lag of $T/2$ seconds between the SUs' transmissions, the probability density function (pdf) of duration of collision ($\tau$) would be as follows:

$$p_\tau(\tau) = \begin{cases} \dfrac{\lambda e^{-\lambda(\frac{T}{2}-\tau)}}{1 - e^{-\lambda\frac{T}{2}}}, & 0 < \tau < \frac{T}{2} \\ 0, & \tau > \frac{T}{2} \end{cases} \tag{9}$$

Collision duration may be longer than $T_f/2$ if there is no NACK at the end of a SU transmission frame. This may occur when the channel between PU transmitter to one or both of SUs' receivers has been in deep fade for at least the collision duration. This outage event in a Rayleigh flat fading channel, is a function of maximum Doppler frequency and average fade duration which can be found from the equations (13-15) in [11]. The probability density function of collision duration more than $T_f/2$ is given in [7, eq. (9)]

### C. Throughput

When PU is active, our proposed system is operating in the CS mode and none of the SUs transmits. However, in case of a misdetection, they may start transmitting while the PU is on. Since the probability of misdetection in the CS mode is very low, we ignore such rare possibility. On the other hand, when PU is idle and the system is in the FDTR mode with imperfect SIS capability, SUs transmit in FD mode with total throughput $R_0 = 2\log(1 + \frac{SNR_S}{1+\beta})$ during collision free intervals, and with degraded throughput $R_1 = 2\log(1 + \frac{SNR_S}{1+SNR_p+\beta})$ during collision interval ($SNR_S$ is the signal to noise ratio of one SU, and $SNR_p$ is that of PU at SUs' receiver). If any false alarm is announced in between, throughput will be zero during subsequent sensing interval. Denoting the transmission time without collision in a frame by $\theta$, and collision duration by $\tau$, the average throughput of a frame would be:

$$\bar{R}_T = P_\tau\left(R_0\frac{\bar{\theta}}{T} + R_1\frac{\bar{\tau}}{T}\right) + R_0(1 - P_\tau). \tag{10}$$

where $\bar{\tau}$ and $\bar{\theta}$ are, respectively, the average values of non-collision and collision durations in asynchronous FD and are derived as follows:

$$\bar{\tau} = \frac{\lambda T + e^{-\lambda T} - 1}{2\lambda(1 - e^{-\lambda T})}, \quad \bar{\theta} = \frac{1 + \lambda T - (2\lambda T + 1)e^{-\lambda T}}{2\lambda(1 - e^{-\lambda T})}. \tag{11}$$

First two terms in (10) correspond to the case when the frame encounters collision, and the second term to the frame with no collision. In order to calculate the average throughput, we need to consider the effect of false alarms on it. As we have two types of false alarms, we need to differentiate between them. False alarm effects on the first frame following the CS mode differently from other frames within the FDTR mode. When the primary signal disappears, the CR network would detect an opportunity with probability of $1 - P_{f1}$ and will start transmission for at least half of frame duration, or will miss the opportunity for $T_s$ seconds (sensing period) with probability of $P_{f1}$. Within the FDTR mode, if a false alarm occurs (with probability of $P_{f2}$), the system will go into the CS mode for at least one sensing slot (i.e. $T_s$ seconds), otherwise will continue transmission for $T/2$ seconds by the end of which another ACK/NACK signal is issued. Given the above breakdown, we derive the average throughput of the system within the modes of CS and FDTR as follows:

$$R_{avg} \cong 2P(\mathcal{H}_0)\left(1 + \frac{2T_s}{\lambda}(P_{f2} - P_{f1}) - \frac{2T_s}{T}P_{f2}\right)\{P_\tau(R_0\bar{\theta} + R_1\bar{\tau}) + R_0(1 - P_\tau)\}. \tag{12}$$

Having inserted the respective values for $\bar{\tau}$ and $\bar{\theta}$ from (11) and after some manipulation, the average throughput of a CR network with the proposed FD scheme, in asynchronous and synchronous transmission modes, respectively are derived as:

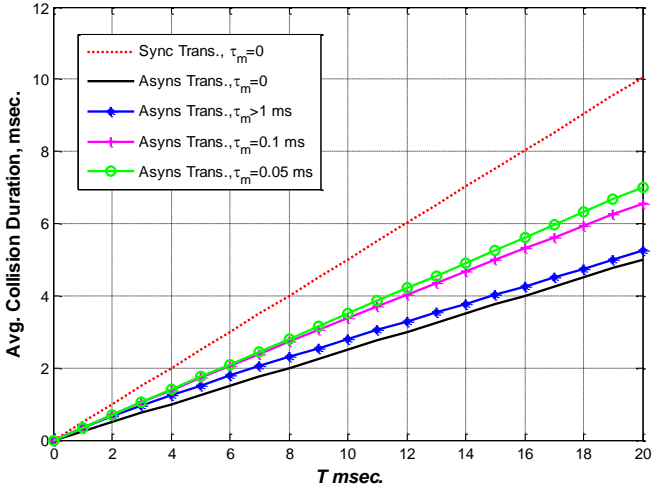
Fig. 3. Average collision duration vs. SU frame duration $T$ for asynchronous and synchronous modes.

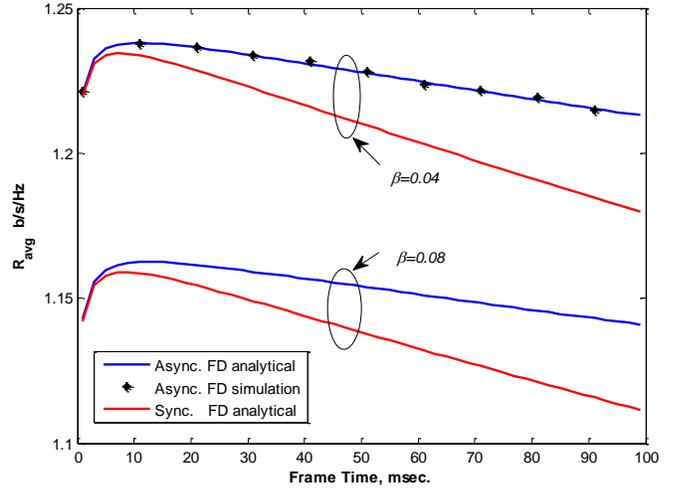
Fig. 4. Normalized average throughput for asynchronous and synchronous modes.

$$R_{Async.} = 2\,P(\mathcal{H}_0)\left(1 + \frac{2T_s}{\lambda}(P_{f2} - P_{f1}) - \frac{2T_s}{T}P_{f2}\right)\left[(R_0 - R_1)\frac{1 - \lambda T - e^{-\lambda T}}{2\lambda T} + R_0\right], \quad (13)$$

$$R_{Sync.} = 2P(\mathcal{H}_0)\left(1 + \frac{2T_s}{\lambda}(P_{f2} - P_{f1}) - \frac{2T_s}{T}P_{f2}\right)\left[(R_0 - R_1)\frac{1 - \lambda T - e^{-\lambda T}}{\lambda T} + R_0\right]. \quad (14)$$

## IV. SIMULATION RESULTS

For simulation purposes we considered a CRN with a pair of SUs and PUs. Here are the parameters used: Activity parameters of the PUs; $\lambda^{-1} = 150$ ms and $\mu^{-1} = 300$ ms, $f_s$= 1MHz, BPSK modulation with convolutional coding, packet size= 1080 bits, CR frame size= 2 packets, SUs' signal to noise ratio= 10 dB, primary SNR at SUs =3 dB, and size of sensing slot $T_s = 1$ ms. We have used the parameters presented in [10, Tab. I] for calculating PER of CRN.

Fig. 3 shows the average collision duration for the synchronous and asynchronous schemes under different fading conditions of the PU-SU channels. When there is no fading, we have minimum average collision duration for the asynchronous mode (which is half of it for synchronous mode). When we have fading, depending on the maximum Doppler frequency ($f_m$) of the channel, the average collision duration varies. For channels with higher $f_m$, outage probability is small and the average collision duration is near to no fade case. But for lower values of $f_m$, the possibility of deep fade durations increases, and this will result in higher average collision durations, as is depicted in this figure.

Fig. 4 shows a comparison between the achievable SUs' throughput in the asynchronous and synchronous modes. We observe that the simulated average throughput is very close to the theoretical one derived in (13). We also observe that this average in the asynchronous mode is slightly more than in synchronous transmission which is due to the shorter collision duration in the asynchronous mode. Moreover, we observe the effect of SIS factor $\beta$ on throughput; poorer SIS capability (higher $\beta$) results in throughput degradation.

## V. CONCLUSIONS

In this paper we proposed a novel scheme incorporating three modes of operation in a CRN. As a result of cooperative sensing in the CS mode, the probability of detection is high enough to guarantee no collision with the primary signal when the PU is in ON state. Based on the decision made in the CS mode through DTD scheme, the system would enter into the conservative FDTS mode, or bidirectional FDTR mode. The asynchronous transmission of SUs in FDTR mode decreased the probability of collision and the average collision duration. We derived an analytical closed form for the probability of collision, the probability density function of collision duration, and the average throughput for the proposed scheme, and numerical results validated the analysis and advantages of the proposed scheme over conventional CRN schemes.

## REFERENCES

[1] W. Cheng, Xi Zhang, and H. Zhang, "Full duplex wireless communications for cognitive radio networks", arXiv:1105.0034 [cs.IT], April 2011.

[2] Y. Liao, L. Song, Z. Han and Y. Li, "Full duplex cognitive radio: a new design paradigm for enhancing spectrum usage," IEEE communications magazine, vol. 53, no. 5, pp. 138-145, May 2015.

[3] L. T. Tan and L. B. Le, "Distributed MAC protocol design for full-duplex cognitive radio networks," in Proc. Of the IEEE Global Communications Conf. (GLOBECOM), pp. 1-6, June 2015.

[4] W. Cheng, Xi Zhang, and H. Zhang, "Full-duplex spectrum-sensing and MAC protocol for multichannel non time-slotted cognitive radio networks", IEEE Journal on Selected Areas in Communications, vol. 33, no. 5, pp. 820 – 831, May 2015.

[5] S. Huang, X. Liu, and Z. Ding, "Optimal sensing-transmission structure for dynamic spectrum access", in Proc. of the IEEE Infocom'09 Conf., pp. 2295–2303, April 2009.

[6] W. Afifi, Ml. Krunz, "Adaptive transmission-reception-sensing strategy for cognitive radios with full-duplex capabilities", in Proc. of the IEEE International Symposium on Dynamic Spectrum Access Networks (DYSPAN), 2014, pp 149-160, April 2014.

[7] V. Towhidlou, and M. Shikh-Bahaei, "Asynchronous full duplex cognitive radio" in Proc. Of the IEEE 84th vehicular technology conference, Fall 2016, Montreal, Sept. 2016.

[8] W. Zhang, R. K. Mallik, and K. B. Letaief, "Optimization of cooperative spectrum sensing with energy detection in cognitive radio networks", IEEE Trans. on Wireless Comm., vol. 8, no. 12, pp 5761–5766, Dec 2009.

[9] Y.C. Liang, Y. Zeng, E.C.Y. Peh, and A. T. Hoang, "Sensing-throughput tradeoff for cognitive radio networks", IEEE Trans. on Wireless Comm., vol. 7, no. 4, Apr 2008.

[10] Q. Liu, S. Zhou, and G. B. Giannakis, "Cross-layer combining of adaptive modulation and coding with truncated ARQ over wireless links", IEEE Trans. on Wireless Commun., vol. 3, no. 5, Sept. 2004.

[11] J. Lai, and N. B. Mandayam, "Minimum duration outages in rayleigh fading channels", IEEE Trans. on Comm., vol. 49, no. 10, Oct 2001.

[12] A. Shadmand, K Nehra, and M Shikh-Bahaei, Cross-layer design in dynamic spectrum sharing systems, EURASIP Journal on Wireless Communications and Networking 2010 (1), 458472.

[13] A. Olfat and M Shikh-Bahaei, "Optimum power and rate adaptation with imperfect channel estimation for MQAM in rayleigh flat fading channel", Vehicular Technology Conference, 2005. VTC-2005-Fall. 2005 IEEE 62nd 4, 2468.

[14] A. Kobravi and M. Shikh-Bahaei, "Cross-layer adaptive ARQ and modulation tradeoffs, Personal, IEEE Indoor and Mobile Radio Communications," 2007.

[15] K. Nehra, A. Shadmand, and M. Shikh-Bahaei, "Cross-layer design for interference-limited spectrum sharing systems," IEEE Global Telecommunications Conference (GLOBECOM 2010), 2010, pp. 1-5.

[16] F. Zarringhalam, B. Seyfe, M. Shikh-Bahaei, G. Charbit, and H. Aghvami, "Jointly optimized rate and outer loop power control with single-and multi-user detection", IEEE Transactions on Wireless Communications 8 (1), 186-195.